# A combined approach of Lattice Boltzmann Method and Maxwell-Stefan equation for modeling multi-component diffusion in solid oxide fuel cell

Ju'an Huang [a], Cheng Bao [†a,c], Zeyi Jiang [a,b], Xinxin Zhang [a,b,c]

a. School of Energy and Environmental Engineering, University of Science and Technology Beijing, Beijing, 100083, China

b. Beijing Key Laboratory of Energy Conservation and Emission Reduction for Metallurgical Industry, Beijing, 100083, China

c. Beijing Higher Institution Engineering Research Center of Energy Conservation and Environmental Protection, Beijing, 100083, China

**Abstract**: Lattice Boltzmann models provide better understanding with mesoscopic eyesight on multi-component diffusion than macroscopic models. Based on the kinetic theory and starting from the He-Luo model, the state-of-the-art multi-component diffusion Lattice Boltzmann models have defects of the compressible error and the limitations for velocity and viscosity settings in lattice units. With these respects, a new Lattice Boltzmann model is presented based on the advection-diffusion equation and is coupled with the Maxwell-Stefan equation by relaxation time. Without introducing the pressure term into the advection-diffusion equation, the model avoids the compressible error. Furthermore, the velocities for components are calculated in the Maxwell-Stefan equation and not contained in the equilibrium distribution function, the limitations of the velocity and viscosity settings in lattice units for under-relaxation iterations are reduced. Then a simulation for $H_2$-$H_2O$-Ar ternary mass transport in the porous anode of the solid oxide fuel cell is employed to validate the accuracy of the Lattice Boltzmann model. The concentration overpotentials are calculated accordingly and compared to several published continuum-scale and Lattice Boltzmann computations, among them, our model offers a better consistency with the experimental measurments.

---

[†] Corresponding author. Tel: +86-10-62333682, Fax: +86-10-62329145
E-mail: baocheng@mail.tsinghua.edu.cn

**Keywords:** Lattice Boltzmann method; solid oxide fuel cell; multi-component diffusion; mass transfer; concentration overpotential

# 1. Introduction

Multi-component diffusion processes exist widely in many engineering applications. Diffusion is the transport of material due to concentration gradients or more precisely due to gradients in the chemical potential. Mass transfer is commonly described as diffusional phenomena in the presence of convective motion. [1] The most frequently used transport models are quantified by the Fick's law in the macroscopic scale, where advective fluxes are neglected. Fick's law is valid for the particular case of a dilute species diffusing into a bulk phase in binary diffusion system. Nevertheless, the multi-component system widely exists in experimental investigation, which involves three or more components. Because of the interaction among the components, the diffusion process of species *i* depends on concentrations and fluxes (or concentration gradients) of themselves and all the other species in the mixture. [2] This is a substantial difference between multi-component and binary diffusion processes. To describe the more general mass transfer process, the Maxwell-Stefan (M-S) equation is developed for the transport model with thermodynamic non-idealities and external force fields, which can be converted into the form resembles the Fick's law and introduced into the advection-diffusion equation conveniently. Moreover, the M-S equation picks up the effect of chemical potentials on concentration gradients. It is appropriate for the diffusion of macro- and micro-porous, e.g., catalysts, adsorbents, membranes. [3]

Lattice Boltzmann method (LBM) is an explicit numerical method by nature for researching mesoscopic flow field in submicron scale, which can capture the right physics of flow and diffusion over a wide range of Knudsen numbers compared to the more traditional approaches. [4] The traditional CFD methods need to satisfy the continuity equation and conservation equations of macroscopic properties at each iteration step. In LBM, however, the fluid is replaced by fractious particles. These particles stream along given directions (lattice links) and collide at the lattice sites. Furthermore, LBM can handle complex phenomena such as moving boundaries (multiphase, solidification, and melting problems), naturally, without a need for face tracing method as it is in the traditional CFD. Hence, LBM can be used for rarefied gas in complex mesoscopic flow domain, which is not applicable for continuity hypothesis. [5] In macroscopic scale, mass transfer problems can be solved by transport models (e.g., Fick's law, M-S equation) combined with advection-diffusion equation. Advection-diffusion, also known as "bulk motion" or "bulk transport," is an effect for multicomponent mass transfer [6] and including both concentrated species (e.g., hydrogen and oxygen) and dilute species (e.g., nitrogen). LBM is a kind of mesoscopic scale simulation method, which in a position to describe the advection-diffusion process in micropores. Therefore, the mass transfer process can also be simulated by transport models combined with this LB model.

Recent years, many LBM variants are designed to model the transport of multiple species on account of the above advantages. Diffusion occurs in the fluid as a result of the random motion of the molecules. The rate of this movement is a function of the fluid temperature, the viscosity and the size (mass) of the particles. According to this kind of thought, many kinetic models are proposed for the diffusion process in LBM derived from Navier-Stokes (N-S) equations as follow, in which the dependent variable is velocity. Shan and Chen [7] proposed a mass transfer simulation for fluids with multiple components and interparticle forces. The macroscopic equations, which govern the motion of each component, is derived through Chapman-Enskog procedure in their work. The accuracy of the binary mixture model is high. Luo and Girimaji [8] developed a two-fluid LB model for binary mixtures, which was derived from the kinetic-theory mixtures model proposed by Sirrovich [9]. In contrast to Shan's model [7], the mutual diffusion and self-diffusion coefficients of this model are independent of the viscosity. The equilibrium distribution function in Luo's model is origin from He-Luo model [10]. McCracken and Abraham [11] extended Luo's model for fluid components with different molecular weights. Asinari [12] developed a binary LB model derived from the continuous kinetic model proposed by Hamel [13] to avoid the complex function of viscosities in Luo's model, in which Hamel's function was calculated for controlling the effective viscosity of the mixture.

By considering self-collision between particles of different species in kinetic LBM-BGK D2Q9 model, Joshi et al. [14] extended Luo's binary model [8] to multi-component model directly by adding more cross-collision terms, which was able to simulate ternary diffusion over a wide range of Knudsen numbers in the non-continuum regime. The viscosities and diffusivities of components can be tuned in this model independently, the molecular weights difference of the components are also considered. Based on this model, Joshi et al. [4] analyzed multi-component gas transport ($H_2$, $N_2$, $H_2O$) in a solid oxide fuel cell (SOFC) anode, of which the porous structure was numerically reconstructed based on SEM (Scanning Electron Microscope) images, to support meso-scale computation in micropores. Asinari [15] also put forward an LB model for multi-component, which recovers the M-S diffusion model and incompressible N-S equations in the continuum limit, without the restriction of the mixture-averaged diffusion approximation. Since this model has only one single relaxation time, however, the viscosities of each component are the same as the viscosity of the whole mixture. To improve the accuracy of Asinari's model in electrolytes, J Zudrop et al. [16] presented an extended multicomponent LB model. This model also recovers momentum and mass transport according to the incompressible N-S equation and Maxwell-Stefan formulation, respectively. In contrast to Asinari's model, every component has their relaxation time for the lattice Boltzmann equation (LBE) in this model.

Joshi's and Zudrop's models are more comprehensive options for multi-component diffusion, which develop from the previous kinetic mass transport LB models. This article makes a review of these LB models in chapter 2. However, there are some drawbacks inherit from the previous models, and they will also be illustrated detailed in chapter 2. We present a new LB model combined with M-S equation for improvement in chapter 3, which is based on the advection-diffusion equation.

## 2. The review of Joshi's and Zudrop's LB models

Chapter 1 reviews the development of LB models for multi-component diffusion. Most widely used one is firmly based upon kinetic theory and resembles He-Luo model, which can be derived to incompressible N-S equations by Chapman-Enskog expansion. To more explicitly explain the improvement of our LB model for multi-component, we first make a simple revisit the details of Joshi's [4] and Zudrop's [16] models mentioned in chapter 1 for better reading.

### 2.1. Joshi's LB model for multi-component diffusion

The discrete lattice Boltzmann evolution equation (LBE) for component $i$ can be written as [4]

$$f_\alpha^i(r + e_\alpha^i \Delta t, t + \Delta t) - f_\alpha^i(r, t) = \Omega_\alpha^i(r, t) \qquad (1)$$

In Eq. (1), $f_\alpha^i$ is the density distribution function of component $i$ associated with the velocity set $e_\alpha^i$ at any spatial location $r$ and time $t$, the particle velocity for component $i$ can be obtained by $u_i = 1/\rho \sum f_\alpha^i e_\alpha^i$, $\Delta t$ is the time increment. $\Omega_\alpha^i$ is the collision term. $\alpha$ is the distribution number and $e_\alpha^i$ is the $\alpha^{\text{th}}$ discrete velocity.

For two-dimensional simulation, the D2Q9 model is widely used, where $e_\alpha^i$ is given by

$$e_\alpha^i = \begin{cases} (0,0) & \alpha = 0 \\ c_i(\cos((\alpha-1)\pi/2), \sin((\alpha-1)\pi/2)) & \alpha = 1-4 \\ \sqrt{2}c_i(\cos((\alpha-5)\pi/2 + \pi/4), \sin((\alpha-5)\pi/2 + \pi/4)) & \alpha = 5-8 \end{cases} \qquad (2)$$

where $c_i$ is the lattice speed for component $i$. For the remaining species, the discrete velocities are obtained by the different lattice speed (DLS) scheme. [11] For convenience, the lightest component can be assumed as component 1, whose lattice speed in lattice units can be defined as $c_1^* = \Delta x^*/\Delta t^* = 1$. The superscript "*" denotes the variables in lattice units, $\Delta x$ is the discrete mesh step, $M_i$ is the molecular weight of component $i$, then the lattice speeds of other components are defined by [17]

$$c_i = c_1 \sqrt{M_1/M_i} \qquad (3)$$

The collision term $\Omega_\alpha^i$ in Eq. (1) for Joshi's model is defined as

$$\Omega_\alpha^i = \Omega_\alpha^{ii} + \sum_{j=1, j \neq i}^{n} \Omega_\alpha^{ij} + F_\alpha^i \Delta t \qquad (4)$$

In Eq. (4), $\Omega_\alpha^{ii}$ is the self-collision term, which is approximated by the BGK (Bhatnagar–Gross–Krook) model of Eq. (5). $\Omega_\alpha^{ij}$ is the cross-collision term estimated by Eq. (6), which represents the effect of collisions between particles of various species and arises only when there is more than one species and the relative velocity between particles of different species is nonzero. $\mathcal{F}_\alpha^i$ represents the effects due to an external acceleration force, which is defined in Eq. (7). [11,14,18]

$$\Omega_\alpha^{ii} = -\frac{1}{\tau_{v,i}}\left[f_\alpha^i(\mathbf{r},t) - f_\alpha^{i(0)}(\mathbf{r},t)\right] \tag{5}$$

$$\Omega_\alpha^{ij} = -\frac{1}{\tau_{D,ij}}\left(\frac{\rho_j}{\rho}\right)\frac{f_\alpha^{i(eq)}}{c_{s,i}^2}(\mathbf{e}_\alpha^i - \mathbf{u}_m)(\mathbf{u}_i - \mathbf{u}_j) \tag{6}$$

$$\mathcal{F}_\alpha^i = w_\alpha \rho_i \frac{\mathbf{e}_\alpha^i \cdot \mathbf{a}_i}{c_{s,i}^2} \tag{7}$$

In Eq. (5), $\tau_i$ is the relaxation time for self-collision which controls the kinematic viscosity of component $i$ via $v_i = c_{s,i}^2(\tau_{v,i} - 1/2)\Delta t$ and $c_{s,i} = c_i/\sqrt{3}$ is the speed of sound. The relationship of the binary diffusion coefficient $D_{ij}$ to $\tau_{D,ij}$ can be determined as follow [19]

$$D_{ij} = \frac{\rho p}{n^2 M_i M_j}\left(\tau_{D,ij} - \frac{1}{2}\right)\Delta t \tag{8}$$

where $\rho$ and $n$ are the mass density ($\rho_i$ is for component $i$) and the number density of the gas mixtures, respectively. $w_\alpha$ in Eq. (7) is the weight factor, which in the D2Q9 model can be defined as

$$w_\alpha = \begin{cases} 4/9 & \alpha = 0 \\ 1/9 & \alpha = 1-4 \\ 1/36 & \alpha = 5-8 \end{cases} \tag{9}$$

The equilibrium distribution function (EDF) $f_\alpha^{i(0)}$ and $f_\alpha^{i(eq)}$ in Eq. (5) and Eq. (6) can be calculated through the extension of He-Luo model [10] as Eq. (10) and Eq. (11), respectively.

$$f_\alpha^{i(0)} = \left[1 + \frac{1}{c_{s,i}^2}(\mathbf{e}_\alpha^i - \mathbf{u}_m) \cdot (\mathbf{u}_i^{eq} - \mathbf{u}_m)\right] f_\alpha^{i(eq)} \tag{10}$$

$$f_\alpha^{i(eq)} = w_\alpha \rho_i \left[1 + \frac{\mathbf{e}_\alpha^i \cdot \mathbf{u}_m}{c_{s,i}^2} + \frac{(\mathbf{e}_\alpha^i \cdot \mathbf{u}_m)^2}{2c_{s,i}^4} - \frac{\mathbf{u}_m^2}{2c_{s,i}^2}\right] \tag{11}$$

The mass concentration $\rho_i$, equilibrium velocity $\mathbf{u}_i^{eq}$ and mass-averaged velocity $\mathbf{u}_m$ for each component are evaluated in Eqs. (12), (13) and (14). [17]

$$\rho_i = \sum_\alpha f_\alpha^i = \sum_\alpha f_\alpha^{i(0)} \tag{12}$$

$$\rho_i \boldsymbol{u}_i^{eq} = \sum_\alpha \boldsymbol{e}_\alpha^i f_\alpha^i = \sum_\alpha \boldsymbol{e}_\alpha^i f_\alpha^{i(0)} \tag{13}$$

$$\rho \boldsymbol{u}_\mathrm{m} = \sum_{i=1}^n \rho_i \boldsymbol{u}_i^{eq}, \quad \rho = \sum_{i=1}^n \rho_i \tag{14}$$

Because He-Luo model is based on the incompressible N-S equation, the LBM variant (a) is only available for incompressible fluid flow. Xu et al. [18] used this model simulated the concentration overpotentials in SOFC electrodes.

## 2.2. Zudrop's LB model for multi-component diffusion

J Zudrop et al. [16] proposed a further extension of the models mentioned above, which applies to the diffusive driving questions. That model recovers momentum and mass transport according to the incompressible Navier-Stokes equation and Maxwell-Stefan formulation, respectively.

The thermodynamic EDF proposed by J Zudrop et al. [16] is shown as follows

$$f_\alpha^{i(eq)} = w_\alpha \left[ \rho_i s_\alpha^i + \frac{\boldsymbol{e}_\alpha \cdot \rho_i \boldsymbol{u}_i^*}{c_s^2} + \frac{\rho_i \cdot (\boldsymbol{e}_\alpha \cdot \boldsymbol{u}_\mathrm{m})^2}{2c_s^4} - \frac{\rho_i u^2}{2c_s^2} \right] \tag{15}$$

where $\rho_i \boldsymbol{u}_i^*$ is defined as

$$\rho_i \boldsymbol{u}_i^* = \rho_i \boldsymbol{u}_i + \sum_j \Gamma_{ij}^{-1} \rho_j \sum_k \chi_k \frac{B_{j,k}}{C} \phi_j (\boldsymbol{u}_k - \boldsymbol{u}_j) \tag{16}$$

Here, M-S equation has been introduced into Eq. (16), in which $B_{jk} = 1/D_{jk}$ is the Maxwell-Stefan resistivity, $\Gamma_{ij}$ is the thermodynamic factor. $C = B\rho_i/p'$ is the ratio of background density to pressure fluctuations. $\phi_j = 1/\kappa_j$. For ideal gas, $\phi_j = 1$.

The collision parameter of this model is defined as

$$\Omega_\alpha^i = \frac{\Delta t}{\frac{1}{\lambda_i} + \frac{\Delta t}{2}} \left[ f_\alpha^i(\boldsymbol{r}, t) - f_\alpha^{i(eq)}(\boldsymbol{r}, t) \right] + \mathcal{F}_\alpha^i \Delta t \tag{17}$$

The collision parameter $\lambda_i$ is defined by $\lambda_i = -1/\tau_{v,i} = BK/p'$, where $B$ denotes a collision frequency, $p'$ denotes an upper limit of the mixture pressure variations, and $K$ denotes the bulk modulus of the liquid mixture measuring the mixture's resistance to uniform compression.

J Zudrop et al. [16] obtained $c_s^2 = K/\rho$, thus $1/C = -c_s^2 \tau_\nu$, in which $\tau_\nu$ is related to the mass-averaged viscosity of the mixture. Remarkably, the form of Eq. (15) is also similar to the EDF of He-Luo model, which recovers mass conservation, the incompressible N-S equation, and the M-S equation. [16]

## 2.3. Shortcomings

The above LB models describe the diffusion process in convection flow. In contrast to the structure to He-Luo model, the term for interactive drag force between each of two components is added in the LBE of Joshi's model and the EDF of Zudrop's model. These LB models have better accuracy than macroscopic models in multi-component diffusion simulation [16,18]. However, there are some shortcomings in which as following:

(1) Difficult to avoid compressible errors.

Joshi's and Zudrop's LB models are designed for describing diffusivity and incompressible fluid flow simultaneously, which resemble and originate from He-Luo model. As a result, they inherit the characteristics of He-Luo model, including some defects. Guo et al. [20] indicated that $\sum f_\alpha^{i(eq)}$ in He-Luo model is not a constant, the continuity equation derived from this model must not meet the incompressible condition from this. The density of the incompressible N-S equation should be a constant, but the D2Q9 LB model cannot meet this demand. It can be explained by the expression of pressure and mass density as $p = c_s^2 \rho$. On the one hand, the density fluctuation of every node in the numerical model represents the pressure distribution, because they are in a simple linear algebraic relationship. On the other hand, the fluid flow is driven by the pressure gradient. Even though the pressure is set as an independent variable in He-Luo model and the derived models, the relation of $p = c_s^2 \rho$ is still existed if $\rho$ is defined. In these EDFs, the pressure and velocity rather than density and momentum behave as independent variables and even take negative values [21], thus the average pressure of the flow must be specified in advance. This compressible effect might lead to some undesirable errors in numerical simulations. In some cases, especially in practical problems, the average pressure is not known or cannot be prescribed precisely. [20]

(2) High computation cost for setting appropriate velocity $\boldsymbol{u}_i$, viscosity or diffusion coefficient for component $i$ in lattice units to ensure under-relaxation iterations.

    a)    The limitation of maximum velocity $\boldsymbol{u}_i$ for component $i$ in lattice units.

Even in the equimolar counter transport with constant pressure and the total molar flux $\boldsymbol{N}_t = \sum \boldsymbol{N}_i = \boldsymbol{0}$ at the boundary, the mass-averaged velocity $\boldsymbol{u}_m \neq \boldsymbol{0}$ and depends on the mass concentration distribution for all components. Furthermore, the diffusing direction of the reactant and product are different. It is easy to derive that the absolute velocity of the lightest component is greater than the absolute mass-averaged velocity in terms of Eq. (14). For LBEs and EDFs in incompressible models, velocities for flow and each component in every direction must be far less than the speed of sound $c_s$. In practice, the condition of Mach number $Ma = u/c_s < 0.15$ is usually maintained in numerical simulations. [10] Usually, in lattice units, take the case of lightest component 1 in the multi-component system as noted earlier, lattice speed is set as $c_1^* = 1$ an then the speed of sound is $c_{s,1}^* = 1/\sqrt{3}$ according to schemes of D1Q3, D2Q5, D2Q9 and D3Q15 [5]. Thus, lattice velocity should meet the condition as $u_1^* < Ma \cdot c_{s,1}^* < 0.1$. Other components can be treated in the same way. Even though the appropriate initial velocities for all components are set, whose maximum value during the numerical calculation are unknown. Computation will overflow when excessive large velocities exist in lattice units during the iteration. A feasible method for this problem is to ensure $u_i^*$ small enough (e.g., $|u_i^*|<0.005$ and 1/10 to the origin), whereas the characteristic length $l^*$ in lattice units (the number of mesh nodes) will be increased as same times to keep Reynolds number $Re$ and viscosity $v^*$ unchanged ($Re = u^* l^*/v^*$). On the other hand, because the simulation time is determined by characteristic lattice time $t^* = l^*/u^*$, for 2-D model, if $u_i^*$ is decreased to 1/10 of origin, iteration steps and grid number are increased to 10 times and 10×10=100 time respectively, the computation will be increased to 1000 times! A reduction of $v^*$ will lead to the following problem.

b) The under-relaxation factor $\omega$ restrict the lower limit of viscosity or diffusion coefficient in lattice units.

To ensure the stable operation of the simulation, the LBEs are solved with under-relaxation iteration, where the reasonable range of relaxation factor $\omega_i = 1/\tau_i$ in Eq. (5) is $0 < \omega_i < 1$, thus $\tau_i > 1$ for every component. Take component 1 as an example in Joshi's model, kinematic viscosity can be related to relaxation parameter $\tau$ as $v_1^* = 1/3\,(\tau_1 - 1/2)$ and then $v_1 > 1/6$. The lower limit of $v_i$ in Zudrop's model is larger, in which $\omega_i = -\Delta t/(1/\lambda_i + \Delta t/2)$ in Eq. (18), thus $\tau_i = -1/\lambda_i > 3/2$ and then $v_1 > 1/3$. The reasonable range for other components and diffusion coefficient can be obtained similarly. To ensure $\omega_i$ is an under-relaxation factor, the viscosity and diffusivity should be large enough in lattice units. Therefore, the unacceptable high computing cost cannot be avoided in the above shortcoming (2b).

(3) The total mass concentration $\rho$ for mass diffusion is not a constant.

For the incompressible model at specific temperature and pressure, the total mass concentration $\rho$, which is also written as the bulk average density, is assumed as a constant. However, the molecular weights and densities for different components are usually not equal. The total mass concentration (density) distribution varies with the fractions of components in the diffusion process. It is contradictory to the incompressible assumption.

## 3. The LB model for multi-component diffusion based on advection-diffusion equations

In original double distribution function of LBM for convection heat transfer in a plain medium, Guo et al. [22] proposed that the evolution of the temperature field be described by another LB model of a temperature distribution function. Taking into consideration the similarity between heat and mass transfer, the corresponding LB model for the temperature field can be extended to the advection-diffusion process for both heat and mass transfer. [5,22] We propose a new LB model combined with M-S equation thereby, which is based on the advection-diffusion equation. The LBE and EDF, which refer to the LB model for temperature field in ref. [22], are independent to whom for velocity field, unlike the LB models illustrated in chapter 2.

The general advection-diffusion equation for incompressible flow is [23]

$$\frac{\partial \phi}{\partial t} + \nabla \cdot (\boldsymbol{u}\phi) = \nabla \cdot (\Gamma \nabla \phi) + S \tag{18}$$

In Eq. (18), $\phi$ is the dependent parameter, $\Gamma$ is the diffusion coefficient. For mass transfer, $\phi$ stands for species concentration and can be written as $c$, $\Gamma$ represents mass diffusivity $D$, $\boldsymbol{u}$ stands for overall velocity vector of the whole flow field, and $S$ denotes source term as an increment of the quantity $\phi$ or $c$ in the time interval $\partial t$. The calculation method of $\boldsymbol{u}$ for multi-component depends on the type of $\phi$. In mass transfer models, Eq. (18) predicts how diffusion causes the concentration to change with time and can be derived into Fick's second law when the source term $S = 0$, overall velocity $\boldsymbol{u} = \boldsymbol{0}$ and $\varGamma$ is a constant.

To describe the effects of the driving force between the components, the generalized M-S equation can be cast as [3]

$$-\frac{X_i}{RT}\nabla_T\mu_i = \sum_{\substack{j=1\\j\neq i}}^{n}\frac{X_j\boldsymbol{N}_i - X_i\boldsymbol{N}_j}{C_tD_{ij}} = \sum_{\substack{j=1\\j\neq i}}^{n}\frac{X_j\boldsymbol{J}_i - X_i\boldsymbol{J}_j}{C_tD_{ij}}, \quad i = 1,2,\ldots,n \tag{19}$$

where $X_i = C_i/C_t$ is the molar fraction of component $i$, $C_i$ is the molar concentration of component $i$, $C_t$ is the total molar concentration. $\nabla_T\mu_i$ is the chemical potential gradient, which is the driving force for diffusion. $D_{ij}$ is called Maxwell-Stefan binary diffusivity between component $i$ and $j$, which represents the inverse of a drag coefficient. For an ideal gas mixture, the $D_{ij}$ is largely independent of composition (but is functions of temperature and pressure). The relation of molar flux between $\boldsymbol{N}_i$ and $\boldsymbol{J}_i$ in Eq. (19) is

$$\boldsymbol{N}_i = X_iC_t\boldsymbol{u}_i, \quad \boldsymbol{J}_i = X_iC_t(\boldsymbol{u}_i - \boldsymbol{u}_M) = \boldsymbol{N}_i - X_iC_t\boldsymbol{u}_M, \quad i = 1,2,\ldots,n \tag{20}$$

where $\boldsymbol{u}_i$ is the velocity of component $i$ with respect to a laboratory-fixed coordinate reference frame, $\boldsymbol{u}_M$ is the molar-averaged mixture velocity and also can be called as bulk velocity. The definition of total molar flux $\boldsymbol{N}_t$ and the relationship between $\boldsymbol{u}_M$ and $\boldsymbol{N}_t$ can be determined as

$$\boldsymbol{N}_t = \sum_{i=1}^{n}\boldsymbol{N}_i = C_t\boldsymbol{u}_M, \quad \boldsymbol{u}_M = \frac{1}{C_t}\sum_{i=1}^{n}\boldsymbol{N}_i, \quad i = 1,2,\ldots,n \tag{21}$$

The EDF in LBM for advection-diffusion problem [5], which resembles the equilibrium temperature distribution function [22], can be written as

$$f_\alpha^{eq} = w_\alpha\phi\left[1 + \frac{\boldsymbol{e}_\alpha\cdot\boldsymbol{u}}{c_s^2}\right] \tag{22}$$

Under the above background, Eq. (22) is capable of describing the mass transfer problem in Fick's law. If the M-S equation can be generalized to the form of Fick's law, Eq. (22) can be used for describing multi-component mass diffusion. This idea can be proved by following section 3.1.

## 3.1. Evolution equation for multi-component flow based on advection-diffusion equation

Considering driving force term $\nabla_T \mu_i$ in Eq. (19) for diffusion, which is counterbalanced by the friction with all of the other moving species $j$ [24], Fick's first law can be changed to [3]

$$\boldsymbol{N}_i = -C_t D_i \left(\frac{X_i}{RT} \nabla_T \mu_i\right) \tag{23}$$

Substituting Eq. (19) into Eq. (23), we can define an effective diffusivity $D_i$ for component $i$ from Maxwell-Stefan binary diffusivity $D_{ij}$ in Eq. (24).

$$\begin{aligned}\frac{1}{D_i} &= -\frac{C_t}{N_i}\left(\frac{X_i}{RT}\nabla_T\mu_i\right) = \frac{C_t}{N_i}\sum_{\substack{j=1\\j\neq i}}^{n}\frac{X_jN_i - X_iN_j}{C_tD_{ij}} = \sum_{\substack{j=1\\j\neq i}}^{n}\frac{X_jN_i - X_iN_j}{D_{ij}N_i} \\ &= \sum_{\substack{j=1\\j\neq i}}^{n}\frac{X_j}{D_{ij}} - \frac{X_iN_j}{D_{ij}N_i}, \qquad i = 1,2,\dots,n \end{aligned} \tag{24}$$

In Eq. (24), flux vectors are replaced by their norm from their directions in Eq. (24), i.e., $N_i = -v_i/|v_i| |\boldsymbol{N}_i|$, where $v_i$ is the stoichiometric coefficient of component $i$ and the arithmetic sign of $N_i$ depends on the orientation of the chemical reaction.

In traditional LBM with single relaxation time (LBM-SRT), relaxation parameter $\tau$ is set as the constant in collision term from the discrete equation. Without the consideration for the binary diffusion coefficient $D_{ij}$ in Eq. (8), the relaxation parameter for component $i$ can be related to the diffusion coefficient $D_i$ from Eq. (24) for the LBE in our LB model as: $\tau_{D,i} = D_i/(c_{s,i}^2 \Delta t) + 1/2$. However, the effective diffusivity $D_i$ is determined by molar fractions $X_i$ and $X_j$, which are functions of coordinate position and simulation time in the calculation and can be written as $D_i(\boldsymbol{r}, t)$.

Therefore, the countermeasures are put forward in this paper: $\tau_{D,i}$ in Eq. (25) is also set as the function $\tau_{D,i}(\mathbf{r}, t)$, which has identical nodes to $X_i$ and can be written as

$$\tau_{D,i}(\mathbf{r}, t) = \frac{D_i(\mathbf{r}, t)}{c_{s,i}^2 \Delta t} + \frac{1}{2} \tag{25}$$

Consequently, the M-S equation is coupled to the LBE in our model with Eq. (25). The equilibrium distribution function (EDF) based on advection-diffusion model is appropriate for multi-component diffusion, the expression for component $i$ is

$$f_\alpha^{i(eq)} = w_\alpha C_i \left[ 1 + \frac{\mathbf{e}_\alpha^i \cdot \mathbf{u}_M}{c_{s,i}^2} \right] \tag{26}$$

where $\mathbf{e}_\alpha^i$ and $c_{s,i} = c^i/\sqrt{3}$ can be obtained from Eqs. (2) and (3). The molar concentration $C_i$ of component $i$ is selected as the independent variable of Eq. (26), thus the overall velocity is $\mathbf{u}_M$ the molar-averaged mixture velocity for bulk movement, instead of the mass-averaged velocity $\mathbf{u}_m$. Correspondingly, $f_\alpha^i$ is the molar concentration distribution function of component $i$ in Eq. (26).

To contrast Joshi's and Zudrop's LB models, the velocity $\mathbf{u}_i$ for individual components is not needed in Eq. (26). Instead, the molar flux $\mathbf{N}_i$ for all nodes and components is needed in our LB model, which is not consisted in Joshi's and Zudrop's models and can be derived in section 3.2. The effective diffusivity $D_i$ for component $i$ is obtained by the M-S equation in Eq. (24) in terms of $\mathbf{N}_i$ and governs the relaxation time $\tau_{D,i}(\mathbf{r}, t)$ for the discrete evolution equation afterwards. The velocity $\mathbf{u}_i$ of component $i$ is considered, even it is not manifest in our LB model. It is coupled to the M-S equation and can be calculated from $\mathbf{N}_i$ by Eq. (20). The interactions of components are consisted in $D_i$, in which $\mathbf{u}_i$ is also a factor for governing.

All the values in molar-averaged velocity matrix $\mathbf{u}_M$ should be already obtained in Eq. (26), which at the boundaries can be obtained by the total boundary flux $\mathbf{N}_t$ by Eq. (21). If the $\mathbf{N}_t$ at boundaries cannot be ignored, another incompressible LB model for velocity field needs to be selected and the velocity distribution of $\boldsymbol{u}_M$ should be calculated firstly at every iteration of the simulation for Eq. (26). That is to say, two different LB models need to be solved. [5] He-Luo model [10] is widely used in simple models for incompressible flow. To ignore the compressibility error, Guo's model [20] can be selected as the velocity model. Note that the density $\rho$ in these models should be replaced as the total molar concentration $C_t$.

Because the moment velocity of each component is not considered in this EDF, the collision term $\Omega_\alpha^i$ for whom is defined as

$$\Omega_\alpha^i = -\frac{1}{\tau_{D,i}(\boldsymbol{r},t)}\left[f_\alpha^i(\boldsymbol{r},t) - f_\alpha^{i(eq)}(\boldsymbol{r},t)\right] + \mathcal{F}_\alpha^i \Delta t \qquad (27)$$

Eq. (27) no longer contains the self-collision term, which can be substituted in Eq. (1) for the evolution equation. The external force term for advection-diffusion is source term in Eq. (18) for specific collision direction, in which $\mathcal{F}_\alpha^i = w_\alpha S^i$. If it can be ignored, the discrete evolution equation can be written as

$$f_\alpha^i(\boldsymbol{r} + \boldsymbol{e}_\alpha^i \Delta t, t + \Delta t) - f_\alpha^i(\boldsymbol{r},t) = -\frac{1}{\tau_{D,i}(\boldsymbol{r},t)}\left[f_\alpha^i(\boldsymbol{r},t) - f_\alpha^{i(eq)}(\boldsymbol{r},t)\right] \qquad (28)$$

The governing Eq. (18) for advection-diffusion has been derived from Eq. (1) and (27) through Chapman-Enskog procedure by Guo et al. [22] Based on the Chapman-Enskog expansion and Taylor series method, Eq. (18) can also be expanded to Eq. (1) and (27), which is demonstrated by A.A. Mohamad. [5] Compared to the collision terms in Joshi's LB model (Eq. (4)-(7)) and EDF in Zudrop's LB model (Eq. (15)-(16)) from chapter 2, the structures of our LB model (Eq. (23)-(28)) is much more straightforward.

For the equimolar counter transport without advection, the total molar flux $\boldsymbol{N}_t = \boldsymbol{0}$ at the inlet and outlet boundaries, thus the molar-averaged velocity $\boldsymbol{u}_M$ can be ignored at the whole diffusion field in Eq. (26) in terms of Eq. (21). This model is suitable for the diffusion process in electrode reactions [25], which is only driven by the concentration gradient. If temperature changed little, the pressure $p$ and the total molar concentration $C_t$ can be considered as a constant at the whole diffusion field, both sides of Eq. (26) are divided by $C_t$, the equilibrium diffusion distribution function without accounting for temperature variation can be written as

$$f_\alpha^{i(eq)} = w_\alpha X_i \tag{29}$$

In Eq. (29), $f_\alpha^i$ is the molar fraction distribution function of component $i$. Obviously, the velocity is not involved in Eq. (29), the shortcomings (2) and (3) in chapter 2 are completely avoided accordingly.

## 3.2. Calculation of fluxes and velocities

To obtain $D_i$ in Eq. (24), fluxes are need to be known. For computational convenience, M-S equation can be converted to the form resembles Fick's first law with matrix operation, which is called as Generalized Fick's law. The left member in M-S equation from Eq. (19) in terms of the mole fraction gradients by introducing a $(n-1) \times (n-1)$ matrix of thermodynamic factor $[\Gamma]$, which is also mentioned in Eq. (16). [3]

$$\frac{X_i}{RT}\nabla_T \mu_i = \sum_{j=1}^{n-1} \Gamma_{ij} \nabla X_j, \quad \Gamma_{ij} = \delta_{ij} + X_i \frac{\partial \ln \gamma_i}{\partial X_j}, \quad \delta_{ij} = \begin{cases} 1 & (i = j) \\ 0 & (i \neq j) \end{cases}, \tag{30}$$

$$i, j = 1, 2, \ldots, n-1$$

$\delta_{ij}$ is Kronecker delta. For the ideal gas mixture, $\partial \ln \gamma_i / \partial X_j = 0$, then $[\Gamma] = [I]$, where $[I]$ is the identity matrix. With the combination of Eq. (19) and (30), M-S equation can be turned to matrix form as

$$-C_t[\Gamma](\nabla X) = [B][J] \tag{31}$$

According to the derivation taken by the ref. [3], the elements of the matrix $[B]$ can be defined in terms of M-S diffusivity $D_{ij}$ by Eq. (19) as follows:

$$B_{ii} = \frac{X_i}{D_{in}} + \sum_{\substack{k=1 \\ k \neq i}}^{n} \frac{X_k}{D_{ik}}, \qquad B_{ij(i \neq j)} = -X_i \left( \frac{1}{D_{ij}} - \frac{1}{D_{in}} \right), \tag{32}$$

$$i, j = 1, 2, \ldots, n - 1$$

Then a matrix of Fick diffusivities $[D]$ is defined by using $(n-1) \times (n-1)$ matrix notation as

$$[D] = [B]^{-1}[\Gamma] \tag{33}$$

Note that $D_{ii}$ of matrix $[D]$ in Eq. (33) have no physical meaning. However, the Fickian $D_{ii}$ enter directly into the expression for the fluxes, and represent the proportionality constant between the driving force and the diffusion flux for the $i^{th}$ component. Comparing Eq. (31) with Eq. (33), the generalized Fick's law for multi-component mixtures is obtained as

$$(J) = -C_t[D](\nabla X) \tag{34}$$

$(J)$ represents the column vector of $(n-1)$ diffusion fluxes that $(J) = [J_1 \ J_2 \ \cdots \ J_{n-1}]^T$, whose value is relative to the direction of $(\nabla X) = [\nabla X_1 \ \nabla X_2 \ \cdots \ \nabla X_{n-1}]^T$. For 2-dimensional (2-D) discrete model, whose mesh size is $p \times q$, $\nabla X_i$ with position $(x_0, y_0)$ in the x-direction can be expressed as

$$\nabla X_i(x_0, y_0) = \frac{X_i(x_0 + 1, y_0) - X_i(x_0, y_0)}{\Delta x}, \quad i = 1, 2, \ldots, n-1; \ x_0 \tag{35}$$

$$= 1, 2, \ldots, p-1; \ y_0 = 1, 2, \ldots, q$$

The molar flux $N_i$ can be derived from $J_i$ by Eq. (20), and after that, velocities $u_i$ for species 1 to $n$-1 are obtained. Given $u_M$ and $u_i$ for species 1 to $n$-1, the velocity $u_n$ for last component $n$ can be determined by Eq. (20) and Eq. (21).

### 3.3. Boundary conditions

1) Dirichlet (first-type) boundary condition

Take the case of the left boundary for the D2Q9 model, after streaming processes, distribution functions $f_1^i$, $f_5^i$ and $f_8^i$ are unknown. According to the detailed flux balance, if molar concentrations in the left boundary are noted as $C_i^\infty = C_t^\infty X_i^\infty$, $f_1^i$ at the boundary ($x = 1$) can be calculated from

$$f_i^1(1) = w_1 C_i^\infty + w_3 C_i^\infty - f_i^3(1) \tag{36}$$

For Eq. (29), $c_i^\infty$ in Eq. (36) can be replaced as the molar fraction $X_i^\infty$.

2) Neumann (second-type) boundary condition

Comparing to the Dirichlet boundary condition, $C_i^\infty$ and $X_i^\infty$ are unknown in the Neumann boundary condition. Nevertheless, since boundary fluxes $J_i^\infty$ is the known quantity, $\nabla c_i^\infty$ or $\nabla X_i^\infty$ in the boundary can be calculated by the generalized Fick's law in chapter 3.2, where $(\nabla X^\infty) = -C_t^{-1}[D]^{-1}(J^\infty)$. For Eq. (35) in right boundary, $\nabla X_i^\infty(y_0)$ in x-direction can be expressed as

$$\nabla X_i^\infty(y_0) = \frac{X_i^\infty(y_0) - X_i(p, y_0)}{\Delta x}, \quad i = 1,2,\ldots,n-1;\ y_0 = 1,2,\ldots,q \tag{37}$$

Thus $X_i^\infty$ are obtained from Eq. (37), which can be plugged into Eq. (36) for distribution functions.

For solid stationary or moving boundary condition, non-slip condition, or flow-over obstacles, which can be called as periodic boundaries, the bounce-back scheme should be used. [5] This scheme assumes that a particle just reverses its velocity after colliding with the solid boundary, ref. [21] takes the detailed illustration.

## 3.4. Improvements

(1) Our LB model cannot describe the fluid movement driven by the pressure gradient, the pressure term does not exist in Eq. (18) accordingly. Without the consideration of $p = c_s^2 \rho$, the shortcoming (1) in chapter 2 is avoided.

(2) The velocities for individual components $\mathbf{u}_i$ don't have to be calculated in the EDF from

our model. The molar-averaged velocity $u_M$ is the only one to keep below than 0.1 in lattice scale, and whose maximum value can be estimated by the continuity equation. Hence, there is no need to keep the scale of velocity in lattice units very small, the negative effects of shortcomings (2) in chapter 2 are reduced substantially.

(3) Because the independent variable in our LB model is selected as the molar concentration, the overall velocity can be set as $u_M = 0$ for the equimolar counter transport without advection, which is common in the electrode diffusion. Thus, it is no need to determine the scale of velocity between the physical and lattice units, and the shortcomings (2) in chapter 2 are completely avoided. In Joshi's and Zudrop's models, the independent variable is the mass concentration, $u_m \neq 0$ and unknown for calculation in this setting.

(4) The total molar concentration $C_t$ is a constant at a certain temperature and pressure based on the ideal gas equation of state. It will not be changed with the fraction of components. The shortcoming (3) is avoided.

## 4. Simulation and Discussion

### 4.1. Ternary diffusion model

To verify the accuracy of the LB model proposed in this paper, a simulation for concentration overpotentials in anode-supported solid oxide fuel cell (SOFC) was carried out. The geometry is a single-unit with bipolar channels in the one-cell stack, whose parameters and operating conditions are extracted from the experiment data of Yakabe et al. [26]

The model of simulation studied the cell performance of the anode, where the operating temperature was kept to 750℃. The calculation area was measured in the $H_2$-$H_2O$-Ar ternary gas system, where the molar fraction ratio of $H_2/H_2O$ was fixed at 4:1 to keep the open circuit voltage (OCV) constant. The molar fraction of $H_2$ in the fuel gas was modified by the degree of dilution of $H_2/H_2O$ gas with Argon gas. The concentration overpotentials at 0.3, 0.7 and 1.0 A cm$^{-2}$ were measured in specific molar fraction ratio of $H_2/(H_2+H_2O+Ar)$ by Yakabe et al. [26] respectively. In this simulation, the electrochemical reaction was assumed to occur at the anode/electrolyte layer (A/E) interface instantaneously as

$$H_2 + O^{2-} \rightarrow H_2O + 2e^- \tag{38}$$

For the channel and anode in SOFC, the physical structure of cross section is illustrated as

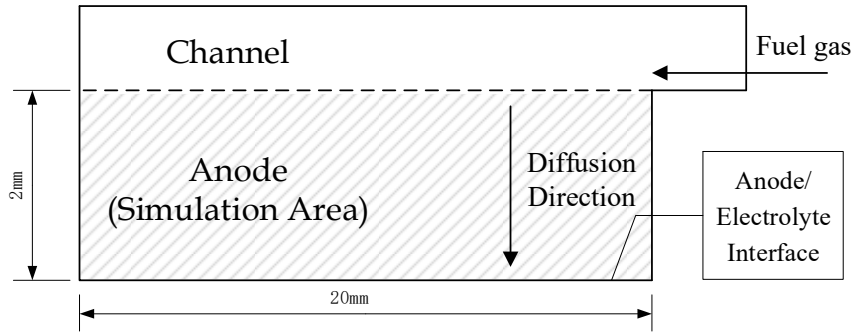

**Fig. 1** The 2-D structure of channel and anode in SOFC.

The D2Q9 scheme was selected for the LB model in this simulation. For the sake of convenience, the cross section of the anode was selected as the simulation area, which was composed of porous media. The dotted line in Fig. 1 was diffusion inlet, and the A/E interface was outlet, whose boundary conditions were Dirichlet and Neumann boundary condition respectively.

At the A/E interface, the flux of species is related to the operating current density ($I$) as [25]

$$N_i|_{A/E} = -v_i\, I/nF \tag{39}$$

where $v_1 = -1$, $v_2 = 1$, $v_3 = 0$ are the stoichiometric coefficients of $H_2$-$H_2O$-Ar in the electrochemical reaction (cf. Eq. (38)), the number of electrons is $n = 2$, $F = 96485\ \text{C mol}^{-1}$ is Faraday constant.

The binary diffusivity $D_{ij}$ between gaseous species $i$ and $j$ can estimated by Fuller et al. [27] expression as

$$D_{ij} = \frac{0.00143 T^{1.75}\sqrt{(1/M_i)+(1/M_j)}}{p\left(V_{F,i}^{1/3}+V_{F,j}^{1/3}\right)^2} \quad [\text{cm}^2\ \text{s}^{-1}] \tag{40}$$

where $T$ is the operating temperature (K), $p$ is operating pressure (bar), $M$ is molar mass (kg kmol$^{-1}$), $V_F$ is the special Fuller diffusion volume. The diffusion in this model is assumed at constant operating temperature and pressure. The values of $M$ and $V_F$ are listed in Table 1. [27]

**Table 1** Gas property data used in the calculation of binary diffusivities.

| Species | Molar mass $M$ (kg kmol$^{-1}$) | Fuller diffusion volume $V_F$ |
|---|---|---|
| H$_2$ | 2.016 | 6.12 |
| H$_2$O | 18.015 | 13.1 |
| Ar | 39.948 | 16.2 |

In SOFC modeling, when applied to diffusional transport within the anode or the cathode, the effective binary diffusivity is usually corrected by accounting for the space-filling aspect, tortuosity and Knudsen number as [25]

$$D_{ij,eff} = \frac{\varepsilon}{2\tau}\left[\frac{1}{1/D_{ij}+1/D_{iM}}+\frac{1}{1/D_{ij}+1/D_{jM}}\right] \tag{41}$$

where

$$D_{iM} = \frac{d_p}{3}\sqrt{\frac{8RT}{\pi M_i}} \tag{42}$$

which keeps the symmetry of the M-S binary diffusivities, i.e. $D_{ij,eff} = D_{ji,eff}$. [25] All the variables from Eq. (41) and (42) are in SI, where $\varepsilon/\tau$ are the porosity-tortuosity ratio, $d_p$ is the average pore diameter [3], $R$ is the gas constant. Total molar concentration can be calculated by the ideal gas equation of state as $c_t = p/RT$. In the simulation, only the tortuosity was treated as a variable parameter, and selected as 4.5 to fit the experimentally measured results at 1.0 A cm$^{-2}$. The values of the main input parameters in this simulation are listed in Table 2. [26]

Table 2 Main experimental and computational parameters of the anode substrate.

| Meaning | Symbol | Value |
| --- | --- | --- |
| Operating temperature (K) | $T$ | 1023.15 |
| Operating pressure (Pa) | $p$ | 1.013×10$^5$ |
| Porosity | $\varepsilon$ | 0.46 |
| Tortuosity | $\tau$ | 4.5 |
| Average pore diameter (μm) | $d_p$ | 2.6 |

Using the Nernst equation, the concentration overpotential $\eta_{conc}$ for ternary (3rd species is inert) fuel system is defined as [25]

$$\eta_{conc} = \frac{RT}{nF} \ln \frac{X_{1,in} X_{2,out}}{X_{1,out} X_{2,in}} \tag{43}$$

In Eq. (43), components 1 and 2 represent H$_2$ and H$_2$O, the subscripts "in" and "out" denote inlet and outlet (A/E interface) boundaries of the anode, which are in Dirichlet and Neumann boundary conditions, respectively.

The size of the anode (simulation area) is 2mm×20mm in diffusion direction as Fig. 1, whose mesh grids are set as 100×1000 in this LB model. The simulation time is set to 2 s, which in lattice unit is designed as 100000 correspondingly. Buckingham π theorem and dimensional analysis can deal with the relationship between physical and lattice units of other parameters. We assume that the chemical reaction occurs at the A/E interface because of the thin thickness anode in the SOFC. Thus, the internal source term $S$ is not existed. Because the temperature and pressure are regarded as the constants approximately, we select Eq. (28) and Eq. (29) for simulation.

For initial settings of the parameters, the fluxes in diffusion direction (x-axis) are assumed to be equal to the boundary fluxes as $N_i(x) = N_i|_{A/E}$. The molar fractions $X_{i,in}$ at inlet boundary is known, hence their initial distributions of the whole anode can be calculated from Eq. (21) and the Generalized Fick's law.

## 4.2. LB model validation

Yakabe et al. [26] and Xu et al. [18] calculated the concentration overpotential $\eta_{conc}$ by macroscopic measurement and Joshi's LB model, which are set as contrast for the simulation data in this paper. The comparison of measured and simulated results are published in Fig. 2-Fig. 4, where the legend "Measured Value" and "LBM" denote the experimental data of Yakabe et al. [26] and the simulation results of our model in this paper, respectively. The vertical axis represents the concentration overpotential ($\eta_{conc}$) minus its basis concentration overpotential ($\eta_{conc0}$), where $\eta_{conc0}$ is obtained at $H_2/(H_2+H_2O+Ar) = 0.8$ (i.e. the molar fraction of Ar is 0). According to the simulation results in this paper, due to the thin thickness of this anode, the variation of flux distributions in the diffusion direction is not obvious that $N_i(x) \approx N_i|_{A/E}$. Therefore, $I(x) \approx I|_{A/E}$ by Eq. (39), the average current density $I_{average} \to I|_{A/E}$.

Figure 2−4 show measured and simulated data points and corresponding fitting curves. In Fig. 2, all the simulation results keep consistent with the experimental data when the average current density is 0.3 A cm$^{-2}$. In Fig. 3, compared to the experimental data, the error is not apparent. The simulation results of Xu et al. [18] and this paper have merit and the shortcoming of each and superior to the simulation results of Yakabe et al. [26] at 0.7 A cm$^{-2}$. In Fig. 4, the deviation of the simulation results from the experimental data are evident at 1.0 A cm$^{-2}$, while the molar fraction of hydrogen ($H_2$) is less than 0.35. Nevertheless, the LB model proposed in this paper has the best

consistency with the experimental data among the listed simulation models.

The simulation model of Yakabe et al. [26] was assumed from Darcy's Law that within the volume containing the distributed resistance. When the Knudsen number is higher than 0.01, Darcy's Law becomes inaccurate. It is the reason that the simulation of Yakabe et al. [26] shows the worst matched results when the Knudsen diffusion is gradually crucial (at low $H_2$ concentrations). The LB model of Xu et al. [18] describes the distribution of velocities for all species that takes into account both diffusion and convection transport. Howbeit Xu et al. [18] used the equivalent porous model, where the pores are not drawn in the mathematical model, the distribution of velocities in this model was the approximate result and not the real one in the actual anode. Under the higher computation burden, Xu's LB model (derived from the N-S equation and has illustrated in chapter 2 (a)) does not work better than our LB model in this simulation.

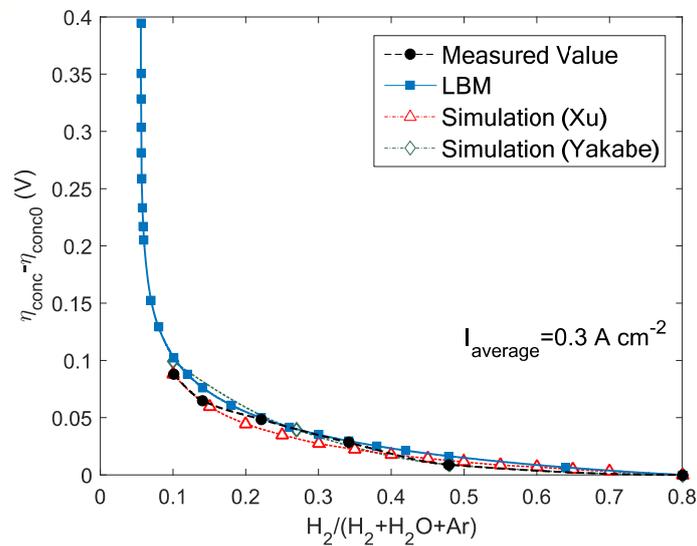

**Fig. 2** Comparison of measured and simulated overpotential ($\eta_{conc}-\eta_{conc0}$) at 0.3 A cm$^{-2}$ in a $H_2$–$H_2O$–Ar system.

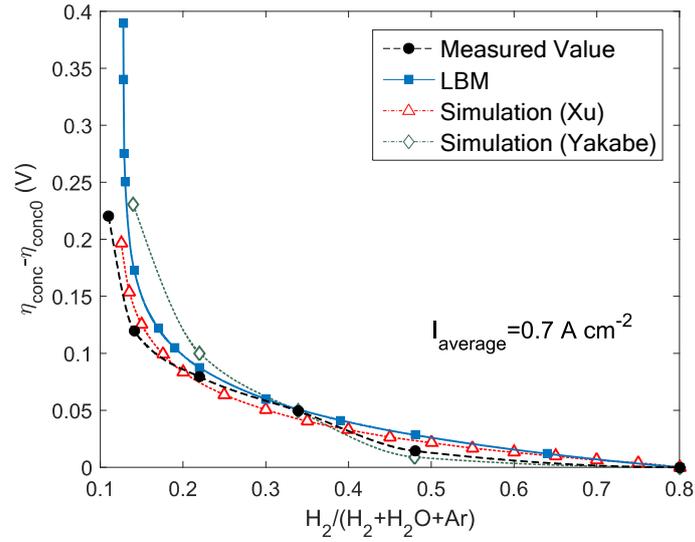

**Fig. 3** Comparison of measured and simulated overpotential ($\eta_{conc}-\eta_{conc0}$) at 0.7 A cm$^{-2}$ in a H$_2$–H$_2$O–Ar system.

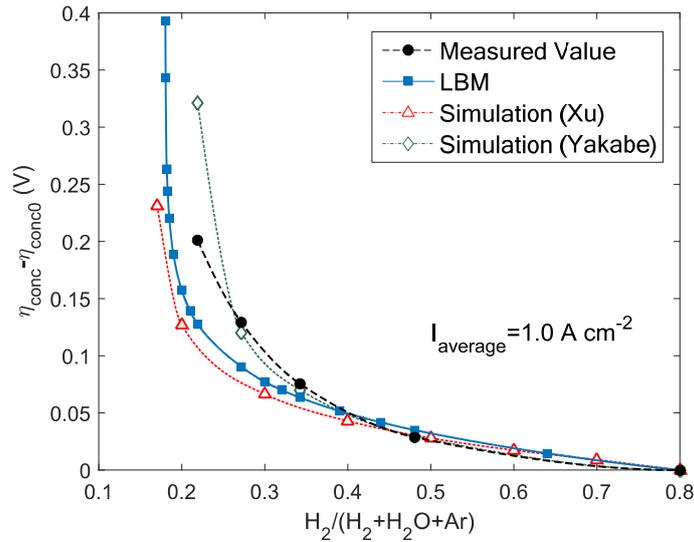

**Fig. 4** Comparison of measured and simulated overpotential ($\eta_{conc}-\eta_{conc0}$) at 1.0 A cm$^{-2}$ in a H$_2$–H$_2$O–Ar system.

## 5. Conclusions

In this article, we proposed a new type of LB model for multi-component diffusion, which was combined with M-S equation and based on the advection-diffusion equation. For the sake of comparison, the LB models based on kinetic theory are analyzed, and after that, their shortcomings are discussed. To verify the reliability of our LB model, the simulation for ternary diffusion in the porous anode of the SOFC is studied, of which results are compared to the measurements of Yakabe et al. [26] The approach in this article comes to some conclusions as follows.

(1) The LB model proposed in this article overcomes some shortcomings of Joshi's [4] and Zudrop's [16] LB models. This model resembles Guo's temperature LB model [22] and is coupled with the M-S equation by relaxation time. Because our model does not contain the pressure term, some undesirable compressible errors in the numerical simulations are avoided. The maximum value of the mass-averaged velocity $u_\mathrm{m}$ and the velocities $u_i$ for each component should be far less than the speed of sound $c_s$ for incompressible LB models, whereas they are difficult to be estimated in Joshi's and Zudrop's models. To ensure them far less than $c_s$ and $0 < \omega_i < 1$ for LBM-SRT simultaneously, the lattice pitch and time-step have to be set very small, which will lead to heavy computation burden. The EDF in our model only contains the molar-averaged velocity $u_\mathrm{M}$, whose maximum value can be estimated by the continuity equation. For the equimolar counter transport without advection, which is common in the electrode diffusion, $u_\mathrm{M} = 0$ and $u_\mathrm{m} \neq 0$, our model is much more straightforward for calculation., The molar concentration is selected as the independent variable in our model, in which total molar concentration $C_t$ is a constant at the specified temperature and pressure. The limitation of the incompressible LB model for a constant density is avoided. In addition, compared to the complicated collision terms in Joshi's model and EDF in Zudrop's model, the structures of our LB model is much simpler.

(2) The concentration overpotentials simulated using D2Q9 LB model proposed in this article have better consistency with the experimental data of Yakabe et al. [26] than that simulated by Yakabe et al. [26] and Xu et al. [18]

(3) The simulation results of our and Xu's LB models have the similar trend, where the deviation of them from the experimental data are apparent at low reactant concentrations and high average current densities.

## Acknowledgments

This work was supported by the Beijing Science and Technology Project [grant number Z181100004518004] and the Fundamental Research Funds for the Central Universities [grant number FRF-GF-17-B31].

## Nomenclature

$C$  molar concentration (mol m$^{-3}$)

$c_s$  speed of sound (m s$^{-1}$)

$D$  diffusion coefficient (m$^2$ s$^{-1}$)

$D_{ij}$  Maxwell diffusivity of component $i$ and $j$  (m$^2$ s$^{-1}$)

$D_{iM}$  Knudsen diffusivity of component $i$  (m$^2$ s$^{-1}$)

$e$  discrete lattice velocity

$F$  Faraday constant, (96485 C mol$^{-1}$)

$\mathcal{F}$  external force term

$f$  lattice Boltzmann distribution function

$I$  operating current density (A cm$^{-2}$)

$\boldsymbol{J}$  diffusion flux for relative velocities (mol m$^{-2}$ s$^{-1}$)

$l$  length (m)

$M$  molecular weight (kg kmol$^{-1}$)

$\boldsymbol{N}$  diffusion flux for absolute velocities (mol m$^{-2}$ s$^{-1}$)

$n$  the number of electrons transferred in the cell reaction or half-reaction

$p$  pressure, (Pa)

$R$  the universal gas constant (8.3145 J K$^{-1}$ mol$^{-1}$)

$S$  source term

$t$  time (s)

$\boldsymbol{u}$  velocity of fluid (m s$^{-1}$)

$v$  stoichiometric coefficient

$w$  weight coefficient

$X$  molar fraction

$x$  coordinate along the direction of anode thickness (diffusion)

$y$  coordinate along the direction of anode length

Greek symbol

$\alpha$  lattice direction

$\Delta t$  time step

$\eta_{conc}$  concentration overpotential (V)

$\nu$  kinematic viscosity (m² s⁻¹)

$\pi$  the ratio of a circle's circumference to its diameter

$\rho$  density / mass concentration (kg m⁻³)

$\tau$  relaxation time

$\Omega$  collision term

Subscripts

$A/E$  interface between anode and electrolyte

$conc$  concentration

$D$  diffusion for mass transfer

$eff$  effective

$t$  total

$i$  component $i$

$j$  component $j$

M  molar average

m  mass average

$in$  inlet

$out$  outlet

$\nu$  viscosity for fluid dynamic

0  basic or specific value

Superscripts

*eq* equilibrium

*i* component *i*